\documentclass[prb,preprint,showpacs,amsmath,amssymb,superscriptaddress]{revtex4}
\usepackage{epsfig}
\usepackage{graphicx}
\usepackage{amsmath,amssymb}

\begin{document}

\title{Experimental observation of the optical spin transfer torque}

\author{P.~N\v{e}mec}
\author{E.~Rozkotov\'a}
\author{N.~Tesa\v{r}ov\'a}
\author{F.~Troj\'anek}
\affiliation{Charles University in Prague, Faculty of Mathematics and Physics, Ke Karlovu 3, 121 16 Prague 2, Czech Republic}
\author{E. De Ranieri}
\affiliation{Hitachi Cambridge Laboratory, Cambridge CB3 0HE, UK}

\author{K.~Olejn{\'{i}}k}
\affiliation{Institute of Physics ASCR, v.v.i., Cukrovarnick\'a 10, 162 53 Praha 6, Czech Republic}

\author{J.~Zemen}
\affiliation{Institute of Physics ASCR, v.v.i., Cukrovarnick\'a 10, 162 53 Praha 6, Czech Republic}

\author{V.~Nov{\'ak}}
\affiliation{Institute of Physics ASCR, v.v.i., Cukrovarnick\'a 10, 162 53 Praha 6, Czech Republic}

\author{M.~Cukr}
\affiliation{Institute of Physics ASCR, v.v.i., Cukrovarnick\'a 10, 162 53 Praha 6, Czech Republic}

\author{P.~Mal\'y}
\affiliation{Charles University in Prague, Faculty of Mathematics and Physics, Ke Karlovu 3, 121 16 Prague 2, Czech Republic}

\author{T.~Jungwirth}
\affiliation{Institute of Physics ASCR, v.v.i., Cukrovarnick\'a 10, 162 53 Praha 6, Czech Republic}
\affiliation{School of Physics and Astronomy, University of Nottingham, Nottingham NG7 2RD, United Kingdom}

\date{\today}

\maketitle
{\bf
The spin transfer torque is a phenomenon in which angular momentum of a spin polarized electrical current entering a ferromagnet is transferred to the magnetization. The effect has opened a new research field of electrically driven magnetization dynamics in magnetic nanostructures and plays an important role in the development of a new generation of memory devices and tunable oscillators.\cite{Ralph:2007_a,Chappert:2007_a} Optical excitations of magnetic systems by laser pulses  have been a separate research field whose aim is to explore magnetization dynamics at short time scales and enable ultrafast spintronic devices.\cite{Kirilyuk:2010_a}  We report the experimental observation of the optical spin transfer torque, predicted theoretically several years ago,\cite{Rossier:2003_a,Nunez:2004_b}  building the bridge between these two fields of spintronics research. In a pump-and-probe optical experiment we measure coherent spin precession in a (Ga,Mn)As ferromagnetic semiconductor excited by circularly polarized laser pulses. During the pump pulse, the spin angular momentum of photo-carriers generated by the absorbed light is transferred to the collective magnetization of the ferromagnet. We interpret the observed optical spin transfer torque and the magnetization precession it triggers on a quantitative microscopic level. Bringing the spin transfer physics into optics introduces a fundamentally distinct  mechanism from the previously reported thermal and non-thermal laser excitations of magnets.  Bringing optics into the field of spin transfer torques decreases by several orders of magnitude the timescales at which these phenomena are explored and utilized.

}

Since the optical spin transfer torque has not yet been reported in the experimental literature we first recall the proposed  physical picture of the phenomenon.\cite{Rossier:2003_a} (Ga,Mn)As ferromagnetic semiconductors utilized in our work are favorable candidates for observing the optical spin transfer torque. The direct-gap GaAs host allows the generation of a high density of photo-carriers, optical selection rules for circularly polarized light yield high degree of spin-polarization of photo-carriers in the direction of the light propagation, and the carrier spins interact with ferromagnetic moments on Mn via exchange coupling. When the ferromagnetic Mn moments are excited, e.g. in the form of coherent magnetization precession, this can be sensitively detected by polarized probe laser pulses due to strong magneto-optical signals in (Ga,Mn)As.\cite{Oiwa:2005_a,Takechi:2007_a,Qi:2007_a,Qi:2009_a,Rozkotova:2008_a,Rozkotova:2008_b,Hashimoto:2008_a,Hashimoto:2008_b}

Coupled precession dynamics of the magnetization orientation $\hat{M}$ and the photo-carrier spin density $\vec{s}$ is governed by the equations,
\begin{eqnarray}
\frac{d\hat{M}}{dt}&=&\frac{J}{\hbar}\hat{M}\times\vec{s}\nonumber\\
\frac{d\vec{s}}{dt}&=&\frac{JS_{Mn}c_{Mn}}{\hbar}\vec{s}\times\hat{M}+P\hat{n}-\frac{\vec{s}}{\tau}\,,
\label{s}
\end{eqnarray}
where $J$ is the carrier - Mn moment exchange coupling constant ($\approx 50$~meVnm$^3$ for holes and $\approx 10$~meVnm$^3$ for electrons in (Ga,Mn)As), $S_{Mn}=5/2$ is the Mn local moment, $c_{Mn}\sim1$~nm$^{-3}$ is the typical Mn moment density, $\hat{M}$ is the unit vector of the magnetization orientation, $P$ is the rate per unit volume at which carrier spins with orientation $\hat{n}$ are optically injected into the ferromagnet, and $\tau$ is the photo-carrier spin relaxation time. In the geometry of our experiments with normal incidence of the laser pulse and in-plane easy-axis of (Ga,Mn)As, the equilibrium $\hat{M}$ is perpendicular to $\hat{n}$. The orientation of $\hat{n}$ is given by the helicity of the circularly polarized laser pulse.

The timescale  of photo-carrier precession due to the exchange field produced by the high density ferromagnetic Mn moments is short; for electrons it is $\sim 2\pi\hbar/JS_{Mn}c_{Mn}\sim$100~fs.  The major source of spin decoherence of the photo-electrons in (Ga,Mn)As is the exchange interaction with fluctuating Mn moments. Microscopic calculations of the corresponding relaxation time give a typical scale of 10's ps.\cite{Rossier:2003_a} The other factor that limits $\tau$ introduced in  Eq.~(\ref{s}) is the photo-electron decay time which is also $\sim$10's~ps, as inferred from reflectivity measurements of the (Ga,Mn)As samples (see Supplementary information). The photo-electron spins therefore precess many times around the exchange field of ferromagnetic Mn moments before  they relax. In the corresponding regime of $\tau JS_{Mn}c_{Mn}/\hbar\gg1$, the steady state spin density of photo-electrons obtained from Eq.~(\ref{s}) by neglecting the last term is,
\begin{equation}
\vec{s}_0\approx\frac{\hbar P}{JS_{Mn}c_{Mn}}(\hat{n}\times\hat{M})\,.
\end{equation}
The steady state spin density is oriented in the plane of the ferromagnetic film and perpendicular to the magnetization vector. The physical interpretation of the result is that the rate of the out-of-plane tilt of $\vec{s}_0$ due to precession around the exchange field produced by the ferromagnetic moments is precisely compensated by the rate of photo-injection of out-of-plane oriented electron spins. This allows the formation of the time independent part of the photo-carrier spin density. From Eq.~(\ref{s}) we see that  $\vec{s}_0$ exerts a torque on the Mn magnetization vector,
\begin{equation}
\frac{dS_{Mn}c_{Mn}\hat{M}}{dt}\approx P\hat{M}\times(\hat{n}\times\hat{M})\,.
\label{stt}
\end{equation}
Eq.~(\ref{stt}) describes the optical spin transfer torque. 

The timescale of the dynamics of $\hat{M}$ due to photo-electrons is $\sim S_{Mn}c_{Mn}/P$ which, for photo-generation rates $P$ in our experiments of $\sim10^{-3}-10^{-2}$~nm$^{-3}$ per $\sim100$~fs, is $\sim 10^2-10^3$ larger than the precession time of the photo-electrons. The fast precessing part of $\vec{s}(t)$ solving Eq.~(\ref{s}) does not, therefore, contribute to the effective torque acting on $\hat{M}$.  Viewed from the perspective of the total angular momentum conservation, the steady state $\vec{s}_0$ mediates a transfer of the entire electron spin injection rate to the rate of change of $\hat{M}$ and no spin angular momentum of the photo-electrons is lost to the environment. The reason for this is that after many precessions, the orientation of the spin of the photo-electron at the moment of relaxation is random, i.e., the sum over all electron contributions to the angular momentum transfer to the environment is zero. This also explains why the optical spin transfer torque in Eq.~(\ref{stt}) is independent of $\tau$. The basic physical picture of the optical spin transfer torque in the  $\tau JS_{Mn}c_{Mn}/\hbar\gg1$ limit is summarized schematically in Fig.~1(a).

The solution of Eq.~(\ref{s}) in the opposite limit of $\tau JS_{Mn}c_{Mn}/\hbar\ll1$, obtained by neglecting the first term, is illustrated in Fig.~1(b). Here the steady state spin density of photo-carriers is out-of-plane and its magnitude scales with $\tau$. The amplitude of the corresponding torque,
\begin{equation}
\frac{dS_{Mn}c_{Mn}\hat{M}}{dt}\approx\frac{\tau JS_{Mn}c_{Mn}}{\hbar} P(\hat{M}\times\hat{n})\,,
\label{stt_tau}
\end{equation}
is then much smaller than the optical spin transfer torque in Eq.~(\ref{stt}). The photo-carrier spins make only a small tilt before relaxing, i.e., most of the angular momentum of the photo-carriers is transferred  to the environment and not to the magnetization vector $\hat{M}$ in this small relaxation time limit. 

The precession time of holes in (Ga,Mn)As is $\sim$10's fs and the spin relaxation time of holes, dominated by the strong spin-orbit coupling, is estimated to $\sim$1-10~fs.\cite{Rossier:2003_a} Since $\tau JS_{Mn}c_{Mn}/\hbar\lesssim1$ for holes, their contribution is better approximated by the weaker torque of Eq.~(\ref{stt_tau}) and we will therefore omit holes in  further discussions. The dominating optical spin transfer torque of photo-electrons described by Eq.~(\ref{stt}) yields a $\sim 0.1-1$ degree out-of-plane tilt of $\hat{M}$ from its equilibrium in-plane orientation for the applied $300$~fs long laser pulses in our experiments. Magnetization precession with these angles are readily detectable  in (Ga,Mn)As.

The experimental observation of the magnetization precession in (Ga,Mn)As excited by the optical spin transfer is presented in Fig.~2(a). The corresponding pump-and-probe measurements were performed on a 20~nm 3.8\% Mn-doped (Ga,Mn)As film grown on a GaAs substrate by low-temperature molecular beam epitaxy. In the experiment, the output of a femtosecond laser is divided into a strong (70~$\mu$Jcm$^{-2}$), 300~fs long pump pulse and a weak delayed probe pulse that are focused to the same spot on the measured sample. The photon energy of 1.64~eV is tuned above the semiconductor band gap  in order to excite magnetization dynamics  by photon absorption.  To observe the optical spin transfer torque we use a circularly polarized pump laser beam.  The pump-induced change of the magneto-optical response of the sample is measured by a time-delayed linearly polarized probe pulse. The magneto-optical signals shown in Fig.~2(a) represent the rotation of the polarization plane of the reflected probe beam. It comprises the signal due to the out-of-plane motion of the magnetization, which is sensed by the polar Kerr effect (PKE), and the signal due to the in-plane component of the ferromagnetic moment, which is sensed by the magnetic linear dichroism (MLD).\cite{Kimel:2005_a} These two contributions can be experimentally separated by their polarization dependence; PKE does not depend on the probe input polarization angle  while MLD is a harmonic function of the input polarization angle. See the Supplementary information for details on sample preparation, experimental techniques, and extensive data on a series of (Ga,Mn)As samples with Mn doping range of 1-14\% .

In the two experiments with circularly polarized pump pulses of opposite helicities, shown in Fig.~2(a), the optically generated photo-carriers have opposite spin orientations, i.e. $P$ in Eq.~(\ref{stt}) is reversed. The initial out-of-plane tilt of $\hat{M}$ due to the optical spin transfer torque and the ensuing phase of the magnetization precession signal should then be also opposite for the two experiments. This key signature of the optical spin transfer torque  is evident in our experimental data shown in Fig.~2(a).

Next we provide a quantitative microscopic analysis of the amplitude and frequency of the oscillating  magneto-optical signals. Since the period of these oscillations (0.4~ns) is much larger than the pump pulse duration, the fast dynamics of the magnetization due to the optical spin transfer torque is reflected only as an initial phase and amplitude of the free precession of the magnetization. The slow free dynamics of the magnetization is governed by  the Landau-Lifshitz-Gilbert torque $\gamma(\hat{M}\times H_{eff})$ with $H_{eff}$ comprising the internal magnetic anisotropy fields and external magnetic field, and by the damping term $\gamma\alpha\hat{M}/|{\bf M}|\times (\hat{M}\times H_{eff})$. Here $\gamma$ is the gyromagnetic ratio and $\alpha$ is the Gilbert damping constant.

We determined the micromagnetic parameters of our (Ga,Mn)As materials by independent magnetization measurements by the superconducting quantum interference device and by magneto-optical ferromagnetic resonance measurements (see the Supplementary information). This allows us to solve the Landau-Lifshitz-Gilbert equation with no fitting parameters. Since the PKE and MLD coefficients were also determined for our (Ga,Mn)As materials from independent static magneto-optical measurements (see Supplementary information), we can transform the calculated time dependent magnetization into the magneto-optical signals and directly compare the calculations with measured data in Fig.~2(a). The initial phase and amplitude is determined in the calculations by the optical spin transfer torque expression~(\ref{stt}). In Fig.~2(b) we show the calculated time evolution of the polar angle $\varphi_M$ and the azimuthal angle $\theta_M$ of the magnetization and in Fig.~2(a) the corresponding theoretical magneto-optical signals (solid lines). In the calculations we set $P\approx1\times10^{-2}$~nm$^{-3}$ per 300~fs. For the applied intensity of the pump laser pulse, this value is comparable to the corresponding photo-carrier injection rate in GaAs. We note that considering the experimental scatter of the measured magneto-optical coefficients and micromagnetic parameters of the studied (Ga,Mn)As, and the relatively small angles of  precession excited in our experiments, the inferred strength of the optical spin transfer torque can vary within a factor of $\sim2$. From our microscopic analysis we conclude that the excitation of magnetization dynamics shown in Fig.~2(a) can be fully ascribed to the optical spin transfer torque. 

In Fig.~3(a) we demonstrate that the previously reported inverse magneto-optical mechanism for exciting  magnetization precession by circularly polarized light is not contributing in our experiments in (Ga,Mn)As. The direct magneto-optical effects represent the dependence of the propagation of polarized light in the sample on the magnetic state of the material. In the inverse magneto-optical effects it is the magnetization which can be modified by the incident light depending on the light polarization. The inverse magneto-optical effects do not require photon absorption and the corresponding theory was derived and the effects were observed in transparent magnetic dielectrics.\cite{Kirilyuk:2010_a} The theory shows that the direct and inverse effects are governed by the same magneto-optical coefficients of the material. In case of circularly polarized light, the relevant phenomena are the inverse Faraday effect in transmission or inverse PKE in reflection. 

To test whether the inverse magneto-optical effect contributes to the excitation of the magnetization in our (Ga,Mn)As sample we tuned the energy of the pump laser beam below the semiconductor band gap to 1.44~eV. Here the optical spin transfer torque is not effective because of the negligible absorption. On the other hand, the below band gap excitation energy is favorable for the inverse magneto-optical effect. The direct PKE is comparable or even larger at 1.44~eV than at 1.64~eV, as shown in Fig.~3(a). Therefore, if active above the band gap at 1.64~eV, the inverse magneto-optical effect should be clearly detectable at 1.44~eV. No magnetization dynamics is, however, excited using the energy of 1.44~eV from which we conclude that the inverse magneto-optical effect is not contributing in our experiments. 

The opposite phase of the measured magneto-optical signals triggered by pump pulses with opposite helicities, shown in Fig.~2(a), suggests that the optical spin transfer torque is also not accompanied by any polarization independent excitation mechanism. Indeed, a linearly polarized pump pulse with any orientation of the polarization is not exciting the magnetization precession, as illustrated in Fig.~3(b). The polarization independent data obtained by summing the signals for the two opposite helicities, and the data corresponding to the linearly polarized pump pulse are nearly identical. They show a non-oscillatory signal which decays on the timescale at which the reflectivity change of the sample due to photo-carriers decay. The signal is therefore of optical and not magnetic origin (for more details see Supplementary information).

Polarization independent excitations of magnetization precession in (Ga,Mn)As have been reported previously by several groups, including us.\cite{Oiwa:2005_a,Takechi:2007_a,Qi:2007_a,Qi:2009_a,Rozkotova:2008_a,Rozkotova:2008_b,Hashimoto:2008_a,Hashimoto:2008_b} The experiments were interpreted in terms of changes of the magnetocrystalline anisotropy by transient increase of temperature or hole-density due to absorbed photons. The sample we study in this work can be also excited by the magnetic anisotropy related mechanism. In the measurements shown in Figs.~2 and 3 we have intentionally suppressed this mechanism in order to highlight our observation of the optical spin transfer torque. We have done this by attaching a PZT piezostressor to the sample which modifies the magnetic anisotropy of the ferromagnetic film due to the differential thermal contraction and allows for an additional {\em in situ} electrical control of the magnetocrystalline anisotropy. The axis of the stressor is aligned at an angle 115$^{\circ}$ from the [100] crystal direction which is close to the easy axis of the bare (Ga,Mn)As epilayer. The induced strain due to the differential thermal contraction is $\sim10^{-3}$ and the additional strain proportional to the applied voltage is $\sim2\times10^{-4}$ for $|U| = 150$~V. These strains can produce sufficient changes of the anisotropy energy. Biasing the stressor makes the direction along its axis magnetically easier for compressive strain, induced by  negative $U$, and the direction perpendicular to the stressor axis magnetically easier for tensile strain, induced by positive $U$.  (More detailed descriptions of the preparation and characterization of the PZT/(Ga,Mn)As hybrid structures are given in Ref.~\onlinecite{Jungwirth:2010_b,Rushforth:2008_a,Ranieri:2008_a} and in the Supplementary information.) 

Applying a large negative piezovoltage of -150~V in the measurements shown in Figs.~2 and 3 we strengthened the magnetic easy axis sufficiently so that the applied laser intensities  did not produce a measurable tilt of the easy axis and, therefore, a measurable magnetization precession. As shown in Figs.~4(a) and (b), the application of a positive piezovoltage allows us to observe the polarization-independent excitation in the same sample. The size of the helicity-dependent signal due to the optical spin transfer torque is comparable to the polarization-independent signal due to the shift of the magnetic easy-axis at the piezovoltage of +150~V. The ratio of the two signals is proportional to the applied piezovoltage, as shown in the Supplementary information. The helicity-dependent signal is observed throughout the series of (Ga,Mn)As materials with different Mn dopings, as we also show in the Supplementary information. Without the attached PZT stressor, the helicity-dependent signal is typically 5-20$\times$weaker than the  polarization-independent signal and was omitted in previous studies of optically excited magnetization precession in (Ga,Mn)As. 

In Fig.~4(c) we illustrate by comparing the measurements at large positive and negative piezovoltages that the helicity-dependent excitation is not sensitive to the variations of the magnetocrystalline anisotropy, as expected for the optical spin transfer torque. Also consistent with the optical spin transfer torque phenomenology, the figure shows that the amplitude of the magnetization oscillations scales with the intensity of the circularly polarized laser pulse. From reflectivity measurements we infer that the photo-carrier density is proportional to the laser pulse intensity in the considered intensity range.

Finally we point out that the measured oscillation periods for the polarization-independent and helicity-dependent excitations are identical. Independent whether the excitation is due  to the shift of the magnetic easy axis or  due to the optical spin transfer torque, the precessing moments are the same in both cases. From studies of the dependence of the precession frequency on internal anisotropy and external magnetic fields it is established that these are the precessing ferromagnetic Mn moments (see Supplementary information and Refs.~\onlinecite{Oiwa:2005_a,Takechi:2007_a,Qi:2007_a,Qi:2009_a,Rozkotova:2008_a,Rozkotova:2008_b,Hashimoto:2008_a,Hashimoto:2008_b}).  

To conclude we have experimentally observed the optical spin transfer torque by measuring the magnetization dynamics in a ferromagnetic semiconductor (Ga,Mn)As induced by a circularly polarized laser pulse. We have interpreted the measured data on a microscopic level. The optical spin transfer torque is a non-thermal mechanism for optical excitation of magnetic systems and we have highlighted the observation of the phenomenon by preparing a sample in which the previously detected non-thermal and thermal mechanisms are ineffective. In our pump-and-probe optical experiments the torque is applied on timescales of 100's fs which is much smaller than the precession period of the magnetization. This represents a qualitatively distinct regime form the studies of the electrically driven spin transfer torque with the typical pulse length of ns's corresponding to many precession periods.

\bibliographystyle{naturemag}

\section*{Corresponding author}
Correspondence and requests for materials should be addressed to Petr N\v{e}mec, nemec@karlov.mff.cuni.cz, Charles University in Prague, Faculty of Mathematics and Physics, Ke Karlovu 3, 121 16 Prague 2, Czech Republic.
\section*{Acknowledgment}
We acknowledge fruitful discussions with A.~V.~Kimel, J.~Sinova, J.~Wunderlich, J.~Fern\'andez-Rossier, and A.~H.~MacDonald, and support  from EU ERC Advanced Grant No. 268066, from the Ministry of Education of the Czech Republic Grants No. LC510 and MSM0021620834, from the Grant Agency of the Czech Republic Grant No. 202/09/H041, from the Charles University in Prague Grant No. SVV-2010-261306 and 443011, and from the Academy of Sciences of the Czech Republic No. AV0Z10100521 and Preamium Academiae.
\section*{Author contributions}
Sample preparation: V.N., M.C., E. R., E. De R.; experiments and data analysis: E.R., N.T., P.N, P. M., K.O, T.J.; data modeling: P.N., F.T.; theory: J.Z., T.J.; writing: T.J., P.N.;  project planning: P.N., T.J.
\section*{Additional information}
Supplementary information accompanies this paper on www.nature.com/naturephysics. Reprints and permissions information is available online at http://npg.nature.com/reprintsandpermissions. Correspondence and requests for materials should be addressed to P.N.

\begin{figure}[h!]
\vspace*{0cm}
\includegraphics[width=.9\columnwidth,angle=0]{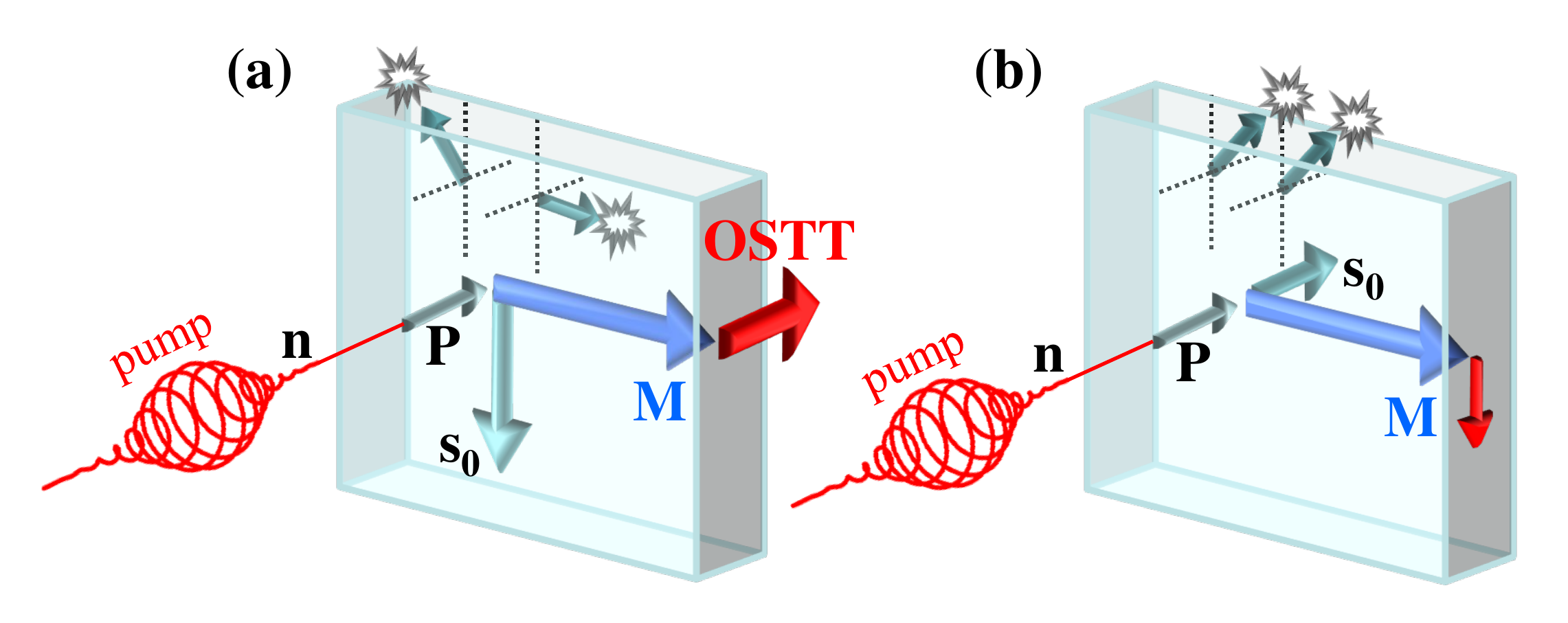}
\vspace*{-0cm}
\caption{(a) Schematic illustration of the optical spin transfer torque in the large spin relaxation time limit. The rate $P$ of the photo-carrier spin injection along light propagation axis $\hat{n}$ (normal to the sample plane) is completely transferred to the optical spin transfer torque (OSTT) acting along the normal to the sample plane on the magnetization $M$ of the ferromagnet. The steady state component of the injected spin density $s_0$ is oriented in the plane of the sample and perpendicular to the in-plane equilibrium magnetization vector. The fast precessing component of the spin of a photo-carrier (small upper arrows) relaxes to the environment at a random orientation, producing a zero net momentum transfer to the environment. This picture applies to photo-electrons in (Ga,Mn)As. (b) A weak torque acting on $M$ produced by photo-carriers with a short 
relaxation time. Most of the spin angular moment is transferred to the environment in this limit. For photo-holes in (Ga,Mn)As this picture is more relevant than the picture of the strong optical spin transfer torque shown in panel~(a).}
\label{fig1}
\end{figure}

\begin{figure}[h!]
\vspace*{0cm}
\includegraphics[width=.55\columnwidth,angle=0]{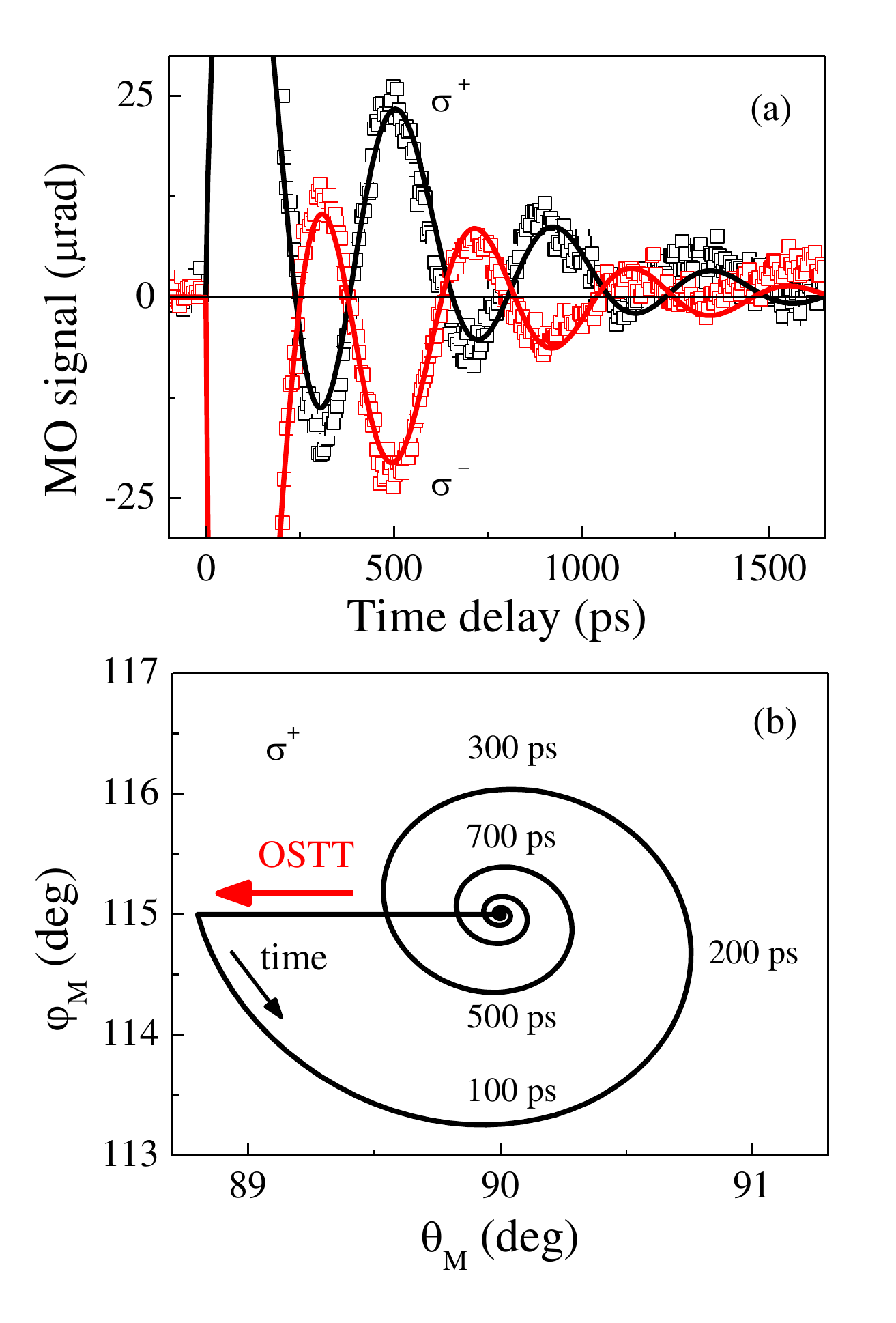}
\vspace*{-0cm}
\caption{Experimental observation of the optical spin transfer torque. (a) Precession of the magnetization induced in (Ga,Mn)As by $\sigma^+$ and $\sigma^-$ circularly polarized pump pulses. Points are the measured rotations of the polarization plane of the reflected linearly polarized probe pulse as a function of the time delay between pump and probe pulses. Lines are the theoretically calculated time-dependent magneto-optical signals. (b) The calculated time evolution of the orientation of the magnetization in the sample, described by the polar angle $\varphi_M$ measured from the [100] axis and the azimuthal angle $\theta_M$ measured from the sample normal [001], induced by the $\sigma^+$ circularly polarized pulse. The orientation of the optical spin transfer torque for $\sigma^+$  polarization is shown in the figure by red arrow (OSTT). For  $\sigma^-$  polarization of the pump pulse, the torque points in the opposite direction. The experiment was performed on the (Ga,Mn)As sample attached to a piezo-stressor at applied bias $U=-150$~V. The probe input polarization orientation is $25\circ$, the  temperature of the measurement 35~K,  the excitation laser intensity 70~$\mu$Jcm$^2$, and the photon energy 1.64~eV.
}
\label{fig2}
\end{figure}
\begin{figure}[h!]
\vspace*{0cm}
\includegraphics[width=.7\columnwidth,angle=0]{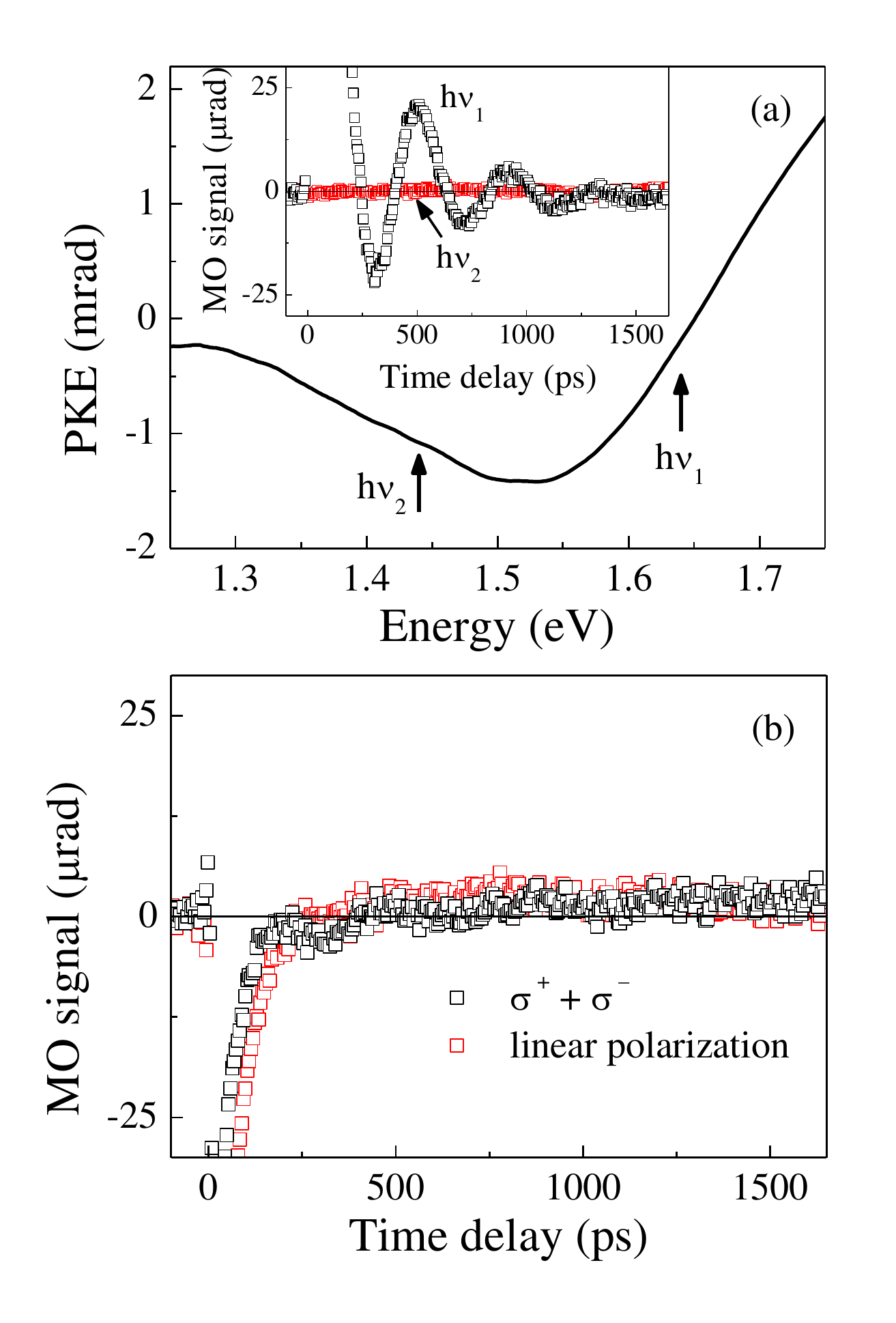}
\vspace*{-0cm}
\caption{(a) Energy dependence of the PKE magneto-optical coefficient. The vertical arrows depict the photon energies $h\nu_1 = 1.64$~eV (above the semiconductor band-gap) and $h\nu_2 = 1.44$~eV (below the semiconductor band-gap) of the circularly polarized pump pulse that were used in the pump-and-probe measurements shown in the inset. Except for the variation of the photon energy of the pump pulse, the experimental conditions of the measurements shown in the inset are the same as in Fig.~2(a). (b) Helicity-independent part [$(\sigma^+ + \sigma^-)/2$] of the signal measured at the pump pulse photon energy 1.64~eV and the pump-and-probe measurement with linearly polarized pump pulse of energy 1.64~eV; all other experimental conditions are the same as in Fig. 2(a).
}
\label{fig3}
\end{figure}

\begin{figure}[h!]
\vspace*{0cm}
\includegraphics[width=.6\columnwidth,angle=0]{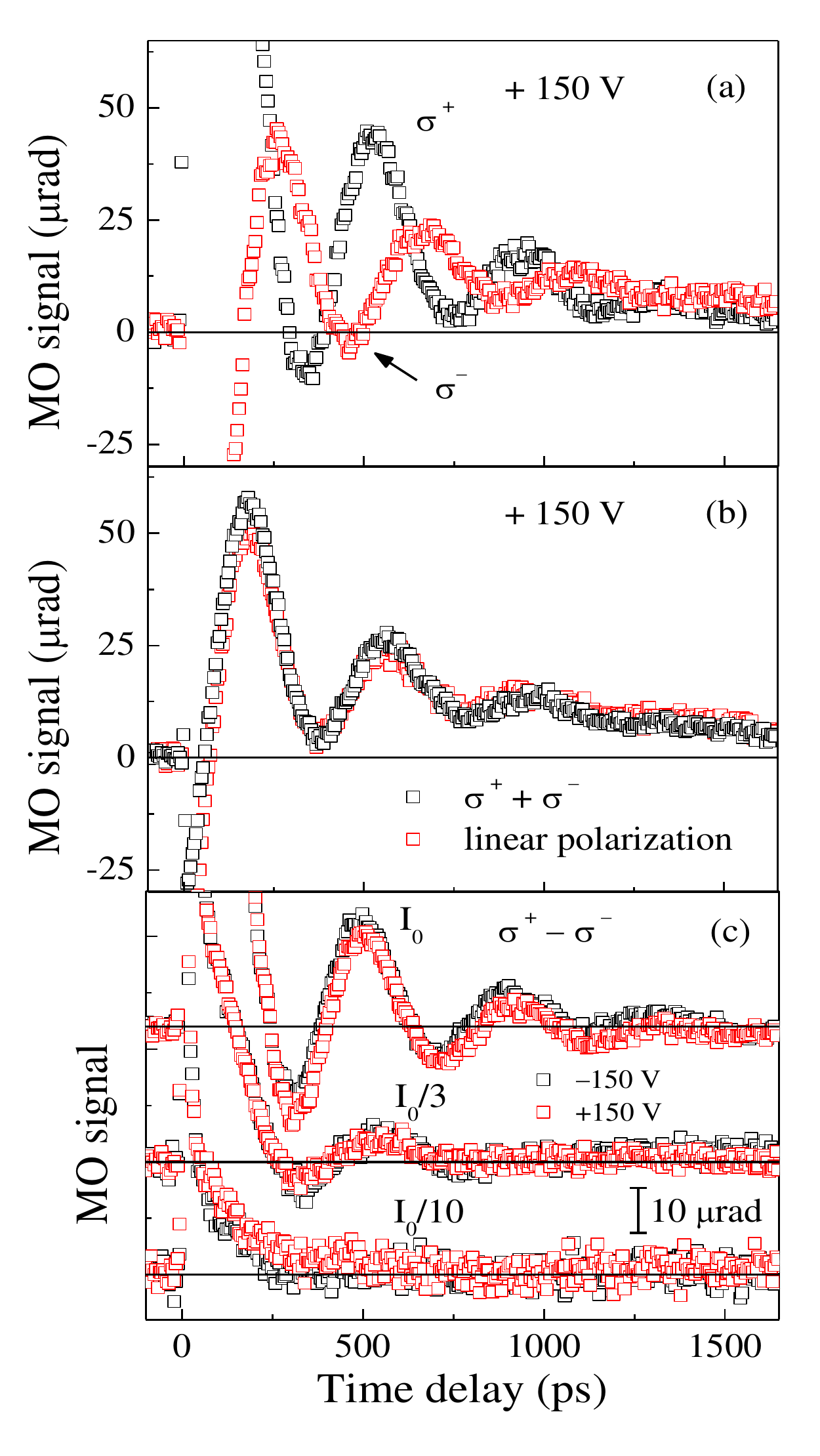}
\vspace*{-0cm}
\caption{(a) Magneto-optical signal measured with $\sigma^+$ and $\sigma^-$ circularly polarized pump pulses and piezo-voltage $U =  +150$~V. (b) Helicity-independent part of the signal shown in panel (a) together with data measured with linearly polarized pump pulses. (c) Intensity dependence ($I_0=70$~$\mu$Jcm$^2$) of the helicity-dependent signal for the piezo-voltages $U =  \pm150$~V. 
}
\label{fig4}
\end{figure}
\end{document}